\documentstyle[prl,aps,epsfig]{revtex}

\twocolumn

\begin{document}
\title{Two-photon Franson-type interference experiments are not tests
  of local realism}

\author{Jan-{\AA}ke Larsson$^{1,*}$, Sven Aerts$^{2,\dagger}$ and Marek
  \.Zukowski$^{3,\ddagger}$}
\address{$^1$Matematiska Institutionen,\\
  Link\"opings Universitet, S-581 83 Link\"oping, Sweden}
\address{$^2$Fundamenten van de Exacte Wetenschappen,\\
  Vrije Universiteit Brussel, Triomflaan 2, 1050 Brussel, Belgium}
\address{$^3$Instytut Fizyki Teoretycznej i Astrofizyki\\
  Uniwersytet Gda\'nski, PL-80-952 Gda\'nsk, Poland} \date{\today}

\maketitle
\begin{abstract}
  We report a local hidden-variable model which reproduces quantum
  predictions for the two-photon interferometric experiment proposed
  by Franson [Phys. Rev. Lett. {\bf 62}, 2205 (1989)].  The model
  works for the ideal case of full visibility and perfect detection
  efficiency.  This result changes the interpretation of a series of
  experiments performed in the current decade.
\end{abstract}
\pacs{03.65.Bz}
\narrowtext

The ingenious two-particle interferometer introduced by Franson
\cite{FRANSON89} is an exciting tool to reveal properties of entangled
states.  In the current decade this device was used in many two-photon
interferometric experiments \cite{FR-EXP}, that beautifully reveal
complementarity between single and two-photon interference.  The
results of the experiments cannot be described using standard methods
involving classical electromagnetic fields \cite{CLASSICAL}.

However, in the original paper, entitled \emph{Bell Inequality for
  Position and Time}, and in many papers that followed, it was claimed
that the experiment constitutes a ``test of local realism involving
time and energy''. Some authors were more sceptical, noting that even
the ideal gedanken model of the experiment involves a postselection
procedure, in which $50\%$ of the events are discarded when computing
the correlation functions \cite{KWIAT}. If all events are taken into
account standard Bell inequalities are not violated. This does not
prove the existence of a local hidden-variable (LHV) model, but merely
states that such a model is not ruled out.

The situation is made even less transparent by similar claims
concerning certain two-photon polarization experiments
\cite{MANDELSHIH} where the problem of discarded events also appears.
This has earlier been treated on equal footing with the problems of
the Franson-type experiments, but a recent analysis in \cite{POPESCU}
of the entire pattern of events in the experiments in
\cite{MANDELSHIH} reestablishes the unconditional violation of local
realism. One could be tempted to adapt the procedure of \cite{POPESCU}
to the Franson experiment, but as will be shown below, this is not
possible.

In short, no one has been able to explicitly show an unconditional
incompatibility of local realistic models of Franson experiments with
the quantum mechanical predictions, but on the other hand, no one has
thus far shown the existence of a LHV model for such predictions
\cite{SANTOS,GARUCCIO}.  Our aim is to resolve this uncertainty about
the interpretation of the Franson interferometry by constructing a LHV
model which fully reproduces the quantum predictions for the ideal
case, i.e for $100\%$ efficient detectors and $100\%$ visibility of
the two-particle interference.

\begin{figure}[htbp]
  \begin{center}
    \psfig{file=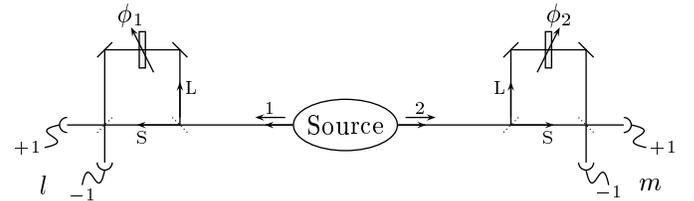}
    \caption{The generic setup of the
      Franson two-photon interference experiment. }
    \label{fig:exp}
  \end{center}
\end{figure}

First, let us describe the main idea behind the Franson type
experiments.  The two-particle interferometer is presented
schematically in Fig.~\ref{fig:exp}. In the actual experiments the
source of the photon pairs was the process of spontaneous parametric
down conversion (PDC) in a nonlinear crystal pumped by a monochromatic
cw laser field. Both the photon traveling to the left as well as the
one traveling to the right were fed into two identical unbalanced
Mach-Zehnder interferometers \cite{MICHELSON}. The difference of the
optical paths in those interferometers, $\Delta L$, satisfies the
relation $cT_{\text{coh}} \ll \Delta L $, where $c$ is the speed of
light and $T_{\text{coh}}$ is the coherence time of the down-converted
photons (it is effectively defined by the filters and the geometry of
the collection of the PDC radiation). Such optical path differences
prohibit any single photon interference. However, two-particle
interference is observable, provided there is no way to know the
actual paths of the photons of a pair which caused two spatially
separated detectors to click.

The down-converted photons have the property that their detection
times (barring retardation effects) are correlated to within their
coherence times \cite{CORRTIME}.  Thus, the two photons either cause
clicks in the two detection stations which are either coincident
(i.e., within coherence times), or one of the clicks is delayed with
respect to the other one by a time difference of the order $\Delta
L/c$.  In the second case, one has the full ``which way'' information
about the process that caused the two clicks (one knows exactly which
photon went via the longer path, and which via the shorter one). In
the first case, however, either both went via the longer arms (denoted
by $L$) or both via the shorter arms ($S$) of the local
interferometers. Therefore, provided the time of emission is unknown
\cite{PULSED}, there is no way to distinguish the two processes, and
thus they interfere.  The relative phase of the quantum amplitudes for
the two processes can be controlled by the phase shifters within the
Mach-Zehnder interferometers, and is equal to their sum.

Formally this can be described in the following simple way. Right
before the exit beamsplitters of the local interferometers
(Fig.~\ref{fig:exp}) the photon state is
\begin{equation}
  \textstyle
  |\psi\rangle_{12} = \frac12 \bigl(|S \rangle_1 + e^{i\phi_1} |L
  \rangle_1\bigr) \bigl(| S\rangle_2 + e^{i\phi_2} |L
\rangle_2\bigr).
\end{equation}
The state vector $| S\rangle_n$ represents the $n$-th photon in the
shorter arm, and $L$ denotes the longer arm. The detectors are located
behind the exit 50--50 beamsplitters, and thus they hide the direct
information about the path of a photon (the indirect information can
still be revealed by the detection times). The part of the state
vector $|\psi\rangle_{12}$ responsible for coincident events (up to
$T_{\text{coh}}$) is given by
\begin{equation}
  \textstyle
  \frac12 \bigl(|S \rangle_1 | S\rangle_2 + e^{i(\phi_1+\phi_2)} |L
  \rangle_1 |L \rangle_2\bigr).
\label{LL}
\end{equation}
As long as the emission times of the pairs of photons are in principle
undefined, the two terms can interfere \cite{PULSED}.  The norm of
this component is $1/\sqrt2$, and thus only half of the events will
belong to this class.  If the exit 50--50 beamsplitters are symmetric,
the probabilities for two-particle processes to give the result
$l=\pm1$ for particle 1 and the result $m=\pm1$ for particle 2, under
specified phase settings, and in coincidence, are
\begin{eqnarray}
  &&P\bigl(l;m(\text{coincidence})|\phi_1,\phi_2\bigr)\nonumber\\
  &&\qquad=\textstyle\frac{1}{8}\bigl(1+lm\cos(\phi_1+\phi_2)\bigr).
  \label{coinc}
\end{eqnarray}

The other terms of $|\psi\rangle_{12}$ cannot lead to coincident
counts, and because the paths taken by the photons is exactly known
there is no interference. The probabilities are
\begin{equation}
  P(l,\text{L};m, \text{E}|{\phi_{1}},{\phi_{2}})
  =P(l,\text{E};m,\text{L}|{\phi_{1}},{\phi_{2}})
  =\textstyle\frac{1}{16},
  \label{noncoinc}
\end{equation}
where $\text{E}$ denotes an {\it earlier} count, and $\text{L}$
denotes a {\it later} count.  No single photon interference is
observed, in other words
\begin{equation}
  \textstyle
  P(l|{\phi_{1}})=P(m|{\phi_{2}})=\frac{1}{2}.
  \label{singles}
\end{equation}

An essential property of any deterministic LHV model for the
experiment is that it should contain the {\it emission} time as one of
the variables describing the experiment. The reason is that the
beamsplitters of, say, the right interferometer may be removed at any
moment of the measurement process. In this case, the photons would be
detected solely by the detector $+1$, and the detection time would
indicate the moment of emission (of course, only up to the coherence
time, but this is not essential).  I.e., there exist an operational
situation in which the emission time can be measured, and therefore it
must be included in the LHV model.  Further, under the same
operational situation, the detections behind the left interferometer
are either coincident with the detections on the right side, or
retarded by $\Delta L/c$.  The LHV model must give predictions for the
local events at one side of the experiment, independent of what
measuring device is used at the other side.  I.e., it must predict
whether a given count on the left side would be coincident (we shall
call this an {\it early} detection) or delayed (a {\it late}
detection) with respect to a count at the right side, when the right
interferometer is dismantled.  From now on we shall assume, just like
in the quantum model, that the average time between emissions is much
longer than all other characteristic times of the experiment.
Otherwise, we assume it to be completely random.

\begin{figure}[htbp]
  \begin{center}
    \psfig{file=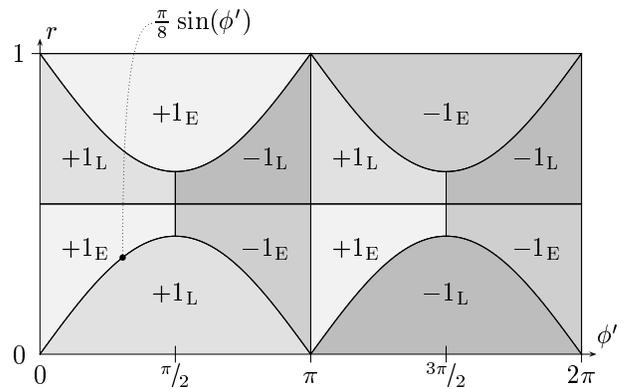} \caption{LHV model for the detection at
      the left station. For a detailed description see the main text.
      The areas labeled $\pm1$ indicate detections at the $\pm1$
      detector, whereas E or L indicate early or late detections. The
      lower curve in the left side of the chart is given by
      $\frac{\pi}{8}\sin{\phi'}$, and the shape of the other curves
      can be easily established by the evident symmetries of the
      chart.}
 \label{fig:det1}
  \end{center}
\end{figure}

Let us now present a LHV model for a single emission at a specific
time for the Franson experiment.  The hidden variables are chosen to
be an angular coordinate $\phi\in[0,2\pi]$ and an additional
coordinate $r\in[0,1]$.  The ensemble of hidden variables is chosen as
that of an even distribution in this rectangle in $(\phi,r)$-space,
but each pair of particles is described by a definite point $(\phi,r)$
in the rectangle, defined at the source at the moment of emission.  At
the left detector station, the measurement result is decided by the
hidden variables $(\phi,r)$ and the local setting $\phi_1$ of the
apparatus. Upon arrival at the detection station, the local variable
$\phi$ is shifted to $\phi'=\phi-\phi_1$, i.e.\ by the current setting
of the local phase shifter.

This shifted value of the angular hidden variable, $\phi'$, together
with $r$, determine the result of the local dichotomic observable
$l=\pm1$ and whether the particle is detected \emph{early} E or
\emph{late} L (Fig.~\ref{fig:det1}).  E.g., if the shifted hidden
variables $(\phi',r)$ end up in an area denoted $+1_{\text{E}}$, the
detector $+1$ fires \emph{early}, if in an area denoted
$-1_{\text{L}}$ the detector $-1$ fires \emph{late}, and so on.

\begin{figure}[htbp]
  \begin{center}
    \psfig{file=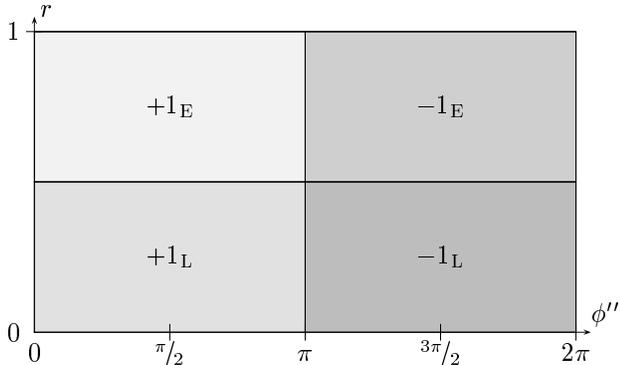} \caption{The measurement result at the
      right station given by the shifted hidden variables.  The
      symbols have the same meaning as in Fig.~\ref{fig:det1}.}
    \label{fig:det2}
  \end{center}
\end{figure}

At the right detector station, a similar procedure is followed.  The
result now depends on $(\phi,r)$ and the local setting $\phi_2$ of the
apparatus. In this case, the shift is to the value
$\phi''=\phi+\phi_2$, and the result is obtained in
Fig.~\ref{fig:det2} in the same manner as before \cite{SYM}.

\begin{figure}[htbp]
  \begin{center}
    \psfig{file=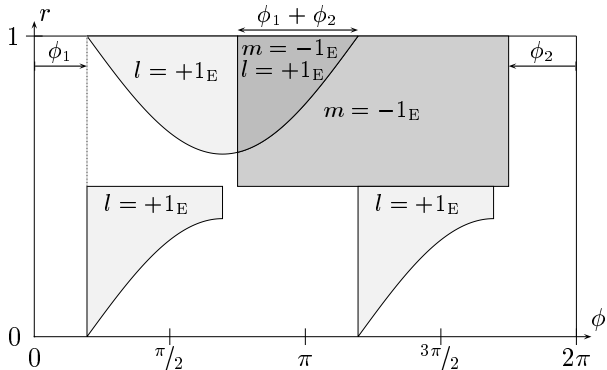} \caption{The shaded regions give the
      values for the initial hidden variables for which
      $l=+1_{\text{E}}$ or $m=-1_{\text{E}}$ are obtained (note that
      $\phi'=\phi-\phi_1$ while $\phi''=\phi+\phi_2$).  The overlap
      region of length $\phi_1+\phi_2$ represents the hidden variables
      for which both $l=+1_{\text{E}}$ and $m=-1_{\text{E}}$ are
      obtained.  }
    \label{fig:turn}
  \end{center}
\end{figure}

The single-particle detection probabilities follow the quantum
predictions as given by (\ref{singles}).  The coincidence
probabilities are determined by interposing Fig.~\ref{fig:det1} and
Fig.~\ref{fig:det2} with the proper shifts.  The probability of having
$l=+1_{\text{E}}$ and $m=-1_{\text{E}}$ simultaneously is the area of
the set indicated in Fig.~\ref{fig:turn} divided by $2\pi$ (the total
area is $2\pi$ whereas the total probability is 1). The probability of
the result $(+1,\text{E};-1,\text{E})$ is easily obtained, and because
of the symmetry of the model, the probability of
$(+1,\text{L};-1,\text{L})$ is the same. Since it is not possible to
distinguish these two results from each other, the probability of
interest is
\begin{eqnarray}
  && P\bigl(+1;-1 (\text{coincidence})|\phi_1,\phi_2\bigr)\nonumber\\
  &&\qquad=P(+1,\text{E};-1,\text{E}|\phi_1,\phi_2)\nonumber\\
  &&\qquad\qquad+ P(+1,\text{L};-1,\text{L}|\phi_1,\phi_2)\nonumber\\
  &&\qquad =\frac{2}{2\pi}
  \int_0^{\phi_1+\phi_2}\frac{\pi}{8}\sin(\phi)d\phi\nonumber \\
  &&\qquad=\frac{1}{8}\bigl(1-\cos{(\phi_1+\phi_2)}\bigr),
\end{eqnarray}
which is equal to the quantum prediction. The other probabilities of
simultaneous detection are obtained in the same manner.  The
probabilities for non-simultaneous detection are also obtained by
integration, but here the symmetry of the model is such that e.g.
\begin{eqnarray}
  &&P(+1,\text{E};-1,\text{L}|\phi_1,\phi_2)\nonumber\\
  &&\qquad=\frac{1}{2\pi} \int_0^{\pi/2}\bigl(
  \frac{1}{2}-\frac{\pi}{8}\sin(\phi) \bigr)d\phi =\frac{1}{16},
\end{eqnarray}
independently of the detector settings, also in accordance with the
quantum predictions.  Finally, if the emission time on one side is
monitored (by removing the interferometer), the counts on the other
side still must follow the same local model, and as it is evident from
the figures 1 and 2 the counts split evenly between early and late
ones.

Let us now move into the conclusions. As the coincident events
constitute only $50\%$ of all events, one might want to dismiss the
whole problem by stating that effectively only around
$\sqrt{50\%}\approx71\%$ events at a single detection station enter
into the Bell analysis, which is much below the usual threshold of
minimum $83\%$ \cite{BOUND}. However, this is also the case in many
other interferometric Bell-type configurations, e.g.\ 
\cite{MANDELSHIH}, but performing a careful analysis it is still
possible to show unconditional violations of local realism for the
quantum predictions describing the expected phenomena in the ideal
case \cite{POPESCU}. The above construction shows that it is not
possible to extend this analysis to Franson-type experiments.

The main conclusion of our work is that all experiments with
Franson-type two-particle interferometers have to be reinterpreted.
They cannot ever serve as demonstrations of violation of local realism
(or, if preferred, violation of a Bell inequality), simply because
there exists a LHV model for the expected quantum predictions in the
ideal case.  We emphasize that this model does not rely on any
imperfections of the actual experiments (like the notorious
detector-efficiency loophole).

One should notice here that majority of the performed quantum
cryptography experiments that involve entanglement are based on the
Franson two-particle interferometry. The original idea of harnessing
quantum entanglement to cryptographic jobs was based on the fact that
security checks can always be performed by testing whether the signals
violate the Bell inequalities \cite{EKERT}.  We have shown that such
violations are only \emph{apparent} for the studied Franson-type
processes.  Therefore, the basis for the application of this type of
phenomena for quantum cryptography has to be carefully re-examined
(compare \cite{CZACHOR}).

Facing our result, one could say that currently only the polarization
entanglement setups are capable to produce true long-distance EPR-Bell
type phenomena. Nevertheless, the Franson two-particle interferometry
remains one of the most beautiful ways to demonstrate the
non-classical nature of light (as the phenomena cannot be described by
any classical field theory). Our \emph{ad hoc} model is important only
in relation to the Bell theorem.

Sven Aerts and Marek \.Zukowski were supported by the Flemish-Polish
Scientific Collaboration Program No. 007. Marek \.Zukowski
acknowledges support of UG Program BW/5400-5-0202-8. Sven Aerts is
supported by the Flemish Institute for the advancement of
Scientific-Technological Research in the Industry (IWT). Jan-{\AA}ke
Larsson acknowledges support from the Swedish Natural Science Research
Council.


\begin{references}
\bibitem{MICHELSON} Some of the laboratory realizations of the
  experiment used two Michelson interferometers. However, such
  devices, due to the inherent loss of counts preclude any link of
  Franson type interferometry with Bell's theorem.  The generic setup
  for Franson type interferometry must involve Mach-Zehnder
  interferometers.
\bibitem{PULSED} This is warranted by the spontaneous nature of the
  PDC emissions, and the continuity of the laser pump field.  An
  interesting feature of the Franson-type interferometry is that the
  PDC source cannot be of pulsed nature (more precisely if the source
  is pulsed, the temporal width of the pulse cannot be shorter than
  $\Delta L/c$).  If there is an unambiguous information about the
  emission time, the two-photon interference cannot occur. Simply,
  comparison of the detection time and the emission time reveals the
  path taken by the particle.
\bibitem{SYM}In the form presented here the model is asymmetric, but
  it may be trivially symmetrized.
\end{references}
\end{document}